\documentclass{Interspeech2024}
\usepackage{multirow}
\usepackage{bigstrut}

\interspeechcameraready

\title{Few-Shot Keyword Spotting from Mixed Speech}
\name[affiliation={1,3}]{Junming}{Yuan}
\name[affiliation={4}]{Ying}{Shi}
\name[affiliation={2*}]{LanTian}{Li}
\name[affiliation={3*}]{Dong}{Wang}
\name[affiliation={1*}]{Askar}{Hamdulla}

\address{
  $^1$School of Computer Science and Technology, Xinjiang University, China\\
  $^2$School of Artiﬁcial Intelligence, Beijing University of Posts and Telecommunications, China \\
  $^3$Center for Speech and Language Technologies, BNRist, Tsinghua University, China \\
  $^4$School of Computer Science and Technology, Harbin Institute of Technology, China 
  \thanks{This work was supported by the National Natural Science Foundation of China (NSFC) under Grants No.62171250/62341607.}
  } 
\email{$^*$Corresponding authors:wangdong99@mails.tsinghua.edu.cn, lilt@bupt.edu.cn, askar@xju.edu.cn}


\begin{document}

\maketitle

\begin{abstract}

Few-shot keyword spotting (KWS) aims to detect unknown keywords with limited training samples. A commonly used approach is the pre-training and fine-tuning framework. While effective in clean conditions, this approach struggles with mixed keyword spotting -- simultaneously detecting multiple keywords blended in an utterance, which is crucial in real-world applications. Previous research has proposed a Mix-Training (MT) approach to solve the problem, however, it has never been tested in the few-shot scenario. In this paper, we investigate the possibility of using MT and other relevant methods to solve the two practical challenges together: few-shot and mixed speech. Experiments conducted on the LibriSpeech and Google Speech Command corpora demonstrate that MT is highly effective on this task when employed in either the pre-training phase or the fine-tuning phase. Moreover, combining SSL-based large-scale pre-training (HuBert) and MT fine-tuning yields very strong results in all the test conditions.

\end{abstract}

\section{Introduction}

Keyword Spotting (KWS), also referred to as spoken term detection, is a task of detecting one or more particular words or phrases from background signals. KWS is a crucial functionality for current intelligent devices to interact via voice. Early research on KWS can be traced back to the 1960s~\cite{teacher1967experimental}. With the emergence of deep learning techniques, researchers have proposed numerous KWS methods based on deep neural nets~\cite{he2017streaming, zhao2020end, zhang2021efficient, liu2022neural, lee2023phonmatchnet}. These studies have elevated KWS performance to a level suitable for large-scale practical applications~\cite{hoy2018alexa, van2022feature}.

A key shortage of the conventional KWS methods is that they were typically designed to detect one or several pre-defined keywords. In practical applications, however, it is often desired to teach the model to learn new keywords with a limited number of examples, a task often named few-shot KWS. Popular approaches to few-shot KWS can be categorized into four types: (1) data augmentation~\cite{gao2020towards, majumdar2020matchboxnet}, (2) Meta-learning~\cite{chen2018investigation, chen2021meta, parnami2022few}, (3) model pre-training~\cite{lin2020training, awasthi2021teaching, mazumder2021few, jung2023metric, rusci2023few}, and (4) utilizing unlabeled data~\cite{seo2021wav2kws, kao2023efficiency, yang2023improving}. Significant progress has been achieved with these approaches. However, almost all these existing studies focus on conditions where only a single speaker appears in the speech signal. This limitation hinders the generalization of few-shot models in scenarios with multiple speakers and both speakers may speak keywords at the same time. Data augmentation (DA)~\cite{seltzer2012acoustic, raju2018data, park2021noisy} is an intuitive strategy, by adding babble noise into the training data. However, DA theoretically cannot address the scenario where the keyword may appear in any of the speech components in the mixed speech. Mixup~\cite{zhang2017mixup} is another promising approach. It mixes speech segments \emph{and} their labels (in one-hot format) of two keywords through linear interpolation, offering improved robustness against mixed speech. However, the training strategy of Mixup makes it unfriendly to keywords with low energy. 

Recently, a new Mix-training (MT)~\cite{shi2023spot} approach was proposed to deal with keywords in mixed speech, allowing the detection of two or more keywords simultaneously speaking. The key design that makes MT superior to Mixup is that the labels of the mixed speech are set to be $k$-hot, to indicate the presence of $k$ keywords, no matter how their energies are strong or weak.
MT effectively reflects the principle of superposition in speech signals and the perceptual experience of the human auditory system. Previous research has demonstrated the superior performance of MT on large-scale keyword datasets, but its performance in few-shot scenarios remains unclear.

Building upon the work in~\cite{shi2023spot}, this paper explores the feasibility of applying MT to few-shot KWS tasks, in particular allowing the detection of simultaneous keywords in mixed speech. 
More specifically, we investigate MT-based pre-training, which extensively uses the MT approach to train a large-scale model that supports detecting speech patterns from mixed speech, and then use this pre-trained model to conduct few-shot fine-tuning. We hypothesise that by exposing the model to a mixed speech during the pre-training stage, the backbone gains the capacity to disentangle the mixed speech and present clean speech patterns to the feature layer. Without this disentanglement and presentation of patterns, the fine-tuning stage struggles to handle, even when employing the MT approach. We will show through our experiments that the MT pre-training can effectively solve the problem, 
especially when MT fine-tuning is employed. 
A key observation from our experiments is that the MT pre-training, although can deal with mixed speech,  does not work well in clean speech conditions. Further investigation shows that the most promising solution is to combine a large-scale SSL pre-trained model like HuBert and MT fine-tuning. This solution yields excellent performance in all the test conditions.



\section{Related work}

\subsection{Few-Shot KWS}

Researchers have attempted various methods to address the few-shot KWS problem. For instance, extensive data augmentation techniques were used in~\cite{gao2020towards, majumdar2020matchboxnet} to increase data diversity. Chen et al.~\cite{chen2018investigation} employed meta-learning methods like MAML to learn better initialization parameters for rapid keyword adaptation. Metric-based meta-learning methods such as Prototypical Networks were also used~\cite{chen2021meta, parnami2022few, snell2017prototypical}. Lin et al.~\cite{lin2020training} pre-trained a multi-head structure as the embedding model using 200M YouTube data, and fine-tuned the model with limited data from new keywords. Mazumder et al.~\cite{mazumder2021few} pre-trained a multi-class, multilingual keyword classification model based on the EfficientNet structure using the Common Voice dataset, and fine-tuned the model to achieve multi-lingual few-shot KWS. Seo et al.~\cite{seo2021wav2kws} employed a Wav2Vec2.0 SSL model as the feature extractor to gain significant generalizability. Kao et al.~\cite{kao2023efficiency} combined SSL with meta-learning to enhance performance in few-shot scenarios.

\subsection{KWS on mixed speech}

Several works have been developed to enhance the robustness of KWS models in noise and other interference scenarios~\cite{yangrobust}. DA is one of the most popular techniques and has been employed in several studies. For instance, Raju et al. and Bonet et al.~\cite{raju2018data, bonet2021speech} applied TV recordings as noise signals to correct keyword speech. However, DA cannot improve KWS from mixed speech. Mixup is another known approach~\cite{zhang2017mixup}. Unlike DA, Mixup combines two keywords through linear interpolation rather than a keyword and an interference speech. According to the mixing strategy, various Mixup algorithms have been proposed, including Manifold mixup ~\cite{verma2019manifold}, CutMix ~\cite{yun2019cutmix}, PixMix ~\cite{hendrycks2022pixmix}, and CoMixup ~\cite{kim2021co}. Shi et al.~\cite{shi2023spot} have demonstrated that Mixup is a vital technique and can deal with KWS from mixed speech to some extent. A shortcoming of Mixup is that it constructs the label of mixed speech by linear interpolation, with the same interpolation ratio when mixing the speech. This linear interpolation means unequal treatment of stronger and weaker speech, making it hard to detect low-energy keywords from mixed speech. The MT approach proposed in~\cite{shi2023spot} solved this problem and has demonstrated a clear advantage on KWS tasks with mixed speech. However, how to deal with the few-shot challenge together with the mixed speech challenge remains an open problem.

\section{Methods}

The research purpose of this study is to employ the MT approach~\cite{shi2023spot} to take the problem of KWS from mixed speech in the few-shot setting. The core problem of few-shot learning is to pre-train a robust model, and the role of the MT approach here is to make the pre-trained model faithfully extract the patterns of all the clean speech components from the mixed signals. We shall firstly revisit the MT method, and then present how the MT-training is implemented.

\subsection{Mix-training strategy}

In short, mix-training is a training strategy for mixed speech. Due to the propositional property of speech waves, a speech signal usually involves sound waves from multiple sources. From the perspective of speech information processing, mixing non-speech waves is not a problem as it does not change the label of the speech signal. However, if the signal involves multiple sources of speech, model training becomes problematic. This has been studied in the name of multi-talker speech recognition, which has achieved rapid progress but the performance is far from perfect~\cite{tripathi2020end}. 

For KWS, mixed speech poses its specific challenge because the existence of multiple speech components makes the detection of target keywords extremely hard, especially for low-energy keywords. Moreover, the strong interference caused by speech mixing poses more risk of false detection. Mix-training was proposed to tackle the mixed speech problem, especially focusing on detecting very weak keywords from strong interference speech~\cite{shi2023spot}.

Briefly, MT consists of three main components: Label union, uniform sampling, and binary cross-entropy (BCE) loss. For label union, MT chose a $k$-shot scheme to reflect the fact that $k$ keywords appear in the signal, no matter how weak they are. Formally, the training speech and its label are constructed as follows:

\vspace{-4mm}
\begin{equation}
     \begin{split}
     x_{mixed} &= \omega_{1}x_{i} + \omega_{2}x_{j}  \\
     y_{mixed} &= y_{i} \oplus y_{j}
     \end{split}
     \label{eq:mt}
\vspace{-1.5mm}
\end{equation}
\noindent where $\omega_{1}$, $\omega_{2}$ are two random variables, and $\oplus$ denotes logical addition. Note that $y_i$ and $y_j$ are one-hot labels.

In MT, a uniform distribution ranging from $(0.1, 0.9)$ is used to sample $\omega_{1}$ and $\omega_{2}$. This uniform sampling means that the energy of the speech components in the mixture can be scaled arbitrarily, but the scaled energy should not be too small, at least above the threshold of human auditory perception. In practice, clean speech is also included in the training process to ensure that the model is exposed to non-mixed data. To avoid clipping, it is often to enforce $\omega_{1} + \omega_{2} = 1$.

Third, MT train models adopt a binary classification architecture trained with the BCE loss. Each binary classifier corresponds to a specific keyword, and the classifier output reflects the probability of the keyword's presence. 

\subsection{MT pre-training}

To employ the MT approach to the few-shot scenario, we follow the pre-training and fine-tuning framework that has been widely used~\cite{mazumder2021few}. In this setting, MT can be employed in both the pre-training phase and the fine-tuning phase. In the pre-training phase, we train a KWS model with a large set of keywords, and a large number of samples for each keyword. This model involves an embedding backbone that projects speech segments of any length to an embedding vector, and a set of linear classifiers, each corresponding to a keyword and predicting if the corresponding keyword exists. In the fine-tuning phase, the classifiers were discarded and a new set of classifiers were reinitialized for the new keywords. During the fine-tuning, the embedding backbone is fixed, and the classifiers are trained with the speech samples for each new keyword. 

Note that in both the pre-training phase and the fine-tuning phase, the training method can be freely selected, including the conventional clean speech training (maybe with data augmentation), Mixup training, or Mix training. By investigating different combinations of these training methods, 
we want to answer the following research questions: (1) Can an MT pre-trained model be more powerful in tackling mixed speech? (2) If the pre-trained model is trained by conventional methods, e.g.,~\cite{mazumder2021few}, is it possible to enhance its capacity to deal with mixed speech by MT fine-tuning?

\section{Experiments}

\subsection{Datasets}


\textbf{Pre-training dataset}: The LibriSpeech\_960 corpus was used in the pre-training phase. Before pre-training the model, we utilized the MFA~\cite{mcauliffe2017montreal} to obtain the word-level alignments for the utterances in Librispeech\_960. Based on the alignment, we chose 5,072 words that involve more than 5 letters and appear more than 100 times. The few-shot target keywords were not in these words. To balance the training, we randomly selected 500 occurrences for high-frequency keywords. The centre position in the speech audio was located for each keyword occurrence, and then a 1s speech segment centred at this position was excerpted as the audio sample of this occurrence. Through this process, we constructed the pre-training dataset that involves about one million (audio, keyword) pairs. To facilitate the pre-training, 1 audio samples were randomly selected for each keyword to form the validation set, and another 1 audio samples per keyword were selected to form the test set. The rest of the samples form the training set.

\noindent \textbf{Fine-tuning dataset}: We used the Google Speech Command (GSC) v2 dataset~\cite{warden2018speech} for the few-shot experiment. The dataset contains 105,829 utterances recorded by 2,618 speakers and includes 35 words. Most of the utterances are 1 second long. The official GSC v2 training set, which involves 35 words, was used to perform fine-tuning. 
Three few-shot conditions were tested: 50-shot, 30-shot and 15-shot. To conduct fine-tuning under these conditions, we constructed a \emph{fine-tuning dataset} for each condition, by randomly sampling 50, 30, and 15 samples respectively for each word in the GSC v2 training set. To observe any randomness, for each condition, we constructed the fine-tuning dataset 5 times by exclusively random sampling and reported the mean and variance of the results on the 5 subsets. The official validation set of GSC v2 was also used for model selection.

\noindent \textbf{Clean test dataset}: The official test set of GSC v2, which involves 10 words, was used for testing in the clean test scenario, where the test sample contains a single speaker.

\noindent \textbf{2-Mix test dataset:} We constructed a 2-mix version for the GSC v2 test set, to support evaluation on mixed speech KWS. We randomly mixed two audio samples of different keywords from the clean test set. 
Note that the same mixing process was also employed to construct a 2-mix version of the test set in the pre-training dataset.

For a clear presentation, the clean test sets are denoted by \emph{Test.clean}, while the 2-mix test sets are denoted by \emph{Test.mixed}. 
The data volume of all these datasets is presented in Table~\ref{tab::dataset}.

\begin{table}[htbp]
  \vspace{-1mm}
  \caption{Datasets used in the experiment.}
  \vspace{-2mm}
  \label{tab::dataset}
   \centering
   \scalebox{0.76}{
		\begin{tabular}{lccccc}
		    \toprule
                 \multirow{2}{*}{} & \multirow{2}{*}{Pre-train} & \multicolumn{3}{c}{Fine-tune} \\
			\cmidrule(r){3-5}
			& & 50-shot & 30-shot & 15-shot   \bigstrut[b]\\
			\midrule
                Train & 1000k & 1.75k & 1.05k & 0.52k \\
                Valid & 5k & 9.90k & 9.90k & 9.90k \\
                Test.clean & 5k & 4.07k & 4.07k & 4.07k \\
                Test.mixed & 5k & 4.07k & 4.07k & 4.07k \\
			\bottomrule
  \end{tabular}}
\end{table}
\vspace{-4mm}

\begin{table}[!htbp]
  \centering
  \caption{Performance of pre-trained models in terms of Top-k ACC(\%) and EER(\%).}
  \vspace{-2mm}
  \label{tab:compare}
  \scalebox{0.8}{
  \begin{tabular}{lcccc}
  \toprule
  \multirow{2}{*}{Pre-train}  &  \multicolumn{2}{c}{Test.clean}  & \multicolumn{2}{c}{Test.mixed}  \\
  \cmidrule(r){2-3} \cmidrule(r){4-5}
             &   Top-1 ACC ($\uparrow$)   & EER ($\downarrow$)   &   Top-2 ACC ($\uparrow$)    & EER ($\downarrow$)   \\			
  \cmidrule(r){1-1} \cmidrule(r){2-3} \cmidrule(r){4-5}
  Clean      &   90.89      & 0.40        &  17.51       &   22.63   \\
  Mixup      &   \textbf{91.84}      & 0.49        &  44.31       &   5.52   \\
  MT         &   89.77      & \textbf{0.39}        &  \textbf{50.75}    &  \textbf{3.87}   \\
  \bottomrule
  \end{tabular}}
  \vspace{-2mm}
\end{table}

\subsection{Pre-training and Fine-tuning}

In the pre-training phase, we followed the EfficientNet recipe\footnote{https://github.com/lukemelas/EfficientNet-PyTorch} and employed the EfficientNet-B0 architecture~\cite{tan2019efficientnet} as the backbone. A linear layer was augmented to predict the posterior of each word in the pre-training set. In the fine-tuning phase, the parameters of the EfficientNet-B0 backbone, obtained from pre-training, were frozen, and two linear layers were initialized and fine-tuned to detect the 10 target keywords in GSC v2.

For both pre-training and fine-tuning, three training strategies were employed: Clean Training (directly training with clean data), Mixup Training, and Mix Training. All the training strategies are based on the BCE loss. There are 9 combinations in total when these training strategies are applied in the pre-training and fine-tuning phases. 

For the Mixup training, the linear interpolation factor was sampled from a Beta(0.2, 0.2) distribution, following the settings in~\cite{zou2023mixupe}. For Mix Training, following the settings in~\cite{shi2023spot}, the scaling factors $\omega_{1}$ and $\omega_{2}$ were separately sampled from a Uniform(0.1, 0.9) distribution, and were then normalized to satisfy $\omega_{1}+\omega_{2}=1$.

The input feature of all the models is 80-dimensional Fbanks derived with a 25ms window size and a 10ms window stride. All the models were trained for 50 epochs with the Adam optimizer, with an initial learning rate set to 1e-3. For each processing, the average of the last 10 checkpoints was delivered to be the final model. The source code will be released later.

Considering the power of large-scale SSL models in feature learning, we compared our pre-trained models with two prevalent SSL models pre-trained by others, 
Wav2Vec 2.0~\cite{baevski2020wav2vec}\footnote{https://huggingface.co/facebook/wav2vec2-base-960h} and 
HuBert~\cite{hsu2021hubert}\footnote{https://huggingface.co/facebook/hubert-base-ls960}. Similarly, during the fine-tuning phase, the parameters of these SSL models were frozen and two linear layers were initialized and fine-tuned.

\begin{table*}[h]
  \centering
  \vspace{-4mm}
  \caption{Results on few-shot keyword spotting}
  \vspace{-2.5mm}
  \label{tab:result}
  \scalebox{0.86}{
    \begin{tabular}{cccccccccccc}
    \toprule
    \multicolumn{12}{c}{\textbf{(a) Top-1 ACC(\%) and EER(\%) with clean test}} \\
    \midrule    
    \multicolumn{3}{c}{Pre-train} & \multicolumn{3}{c}{Fine-tune} & \multicolumn{2}{c}{50-shot} & \multicolumn{2}{c}{30-shot} & \multicolumn{2}{c}{15-shot}   \\
    \cmidrule(r){1-3}    \cmidrule(r){4-6}	 \cmidrule(r){7-8}   \cmidrule(r){9-10}  \cmidrule(r){11-12}
                 Clean & Mixup & MT            & Clean & Mixup & MT            & Top-1 ACC ($\uparrow$) & EER ($\downarrow$)  & Top-1 ACC ($\uparrow$)  & EER ($\downarrow$)  & Top-1 ACC ($\uparrow$) & EER ($\downarrow$) \\
    \cmidrule(r){1-3}    \cmidrule(r){4-6}	 \cmidrule(r){7-8}   \cmidrule(r){9-10}  \cmidrule(r){11-12}
                \checkmark &&& \checkmark &&&  78.12$\pm$1.04   & 8.19$\pm$0.28    & 76.09$\pm$0.73  & 8.71$\pm$0.31  & 73.15$\pm$0.76  & 9.64$\pm$0.36 \\
                \checkmark&&&&\checkmark&&     81.07$\pm$0.20   & 7.00$\pm$0.19    & 79.31$\pm$0.58  & 7.31$\pm$0.36  & 74.98$\pm$0.68  & 9.24$\pm$0.39 \\
                \checkmark&&&&&\checkmark&     81.34$\pm$0.32   & 7.19$\pm$0.04    & 79.84$\pm$0.24  & 7.71$\pm$0.30  & 75.78$\pm$0.74  & 9.16$\pm$0.23 \\
\cmidrule(r){1-3}    \cmidrule(r){4-6}	 \cmidrule(r){7-8}   \cmidrule(r){9-10}  \cmidrule(r){11-12}
                &\checkmark&& \checkmark &&&   84.87$\pm$0.54   & 6.14$\pm$0.23    & 83.07$\pm$0.48  & 6.79$\pm$0.23  & 80.07$\pm$1.30  & 8.15$\pm$0.59 \\
                &\checkmark&&&\checkmark&&     85.64$\pm$0.21   & \textbf{5.68$\pm$0.33}    & 83.93$\pm$0.58  & \textbf{6.46$\pm$0.25}  & 80.62$\pm$1.62  & 7.85$\pm$0.52 \\
                &\checkmark&&&&\checkmark&     \textbf{86.02$\pm$0.51}   & 6.00$\pm$0.19    & \textbf{84.43$\pm$0.61}  & 6.54$\pm$0.31  & \textbf{81.23$\pm$1.80}  & \textbf{7.75$\pm$0.73} \\
\cmidrule(r){1-3}    \cmidrule(r){4-6}	 \cmidrule(r){7-8}   \cmidrule(r){9-10}  \cmidrule(r){11-12}
                &&\checkmark& \checkmark &&&   78.12$\pm$0.41   & 8.22$\pm$0.23    & 75.38$\pm$1.08  & 9.04$\pm$0.35  & 70.03$\pm$1.55  & 11.00$\pm$0.61 \\
                &&\checkmark&&\checkmark&&     80.36$\pm$0.99   & 7.38$\pm$0.20    & 77.29$\pm$0.99  & 8.30$\pm$0.27  & 71.01$\pm$1.12  & 10.65$\pm$0.79 \\
                &&\checkmark&&&\checkmark&     80.63$\pm$0.60   & 7.52$\pm$0.15    & 77.23$\pm$0.57  & 8.63$\pm$0.39  & 71.35$\pm$1.33  & 10.64$\pm$0.73 \\
\cmidrule(r){1-3}    \cmidrule(r){4-6}	 \cmidrule(r){7-8}   \cmidrule(r){9-10}  \cmidrule(r){11-12}
\cmidrule(r){1-3}    \cmidrule(r){4-6}	 \cmidrule(r){7-8}   \cmidrule(r){9-10}  \cmidrule(r){11-12}
          &&  & \checkmark &  &  &            87.65$\pm$0.38  &  5.37$\pm$0.23  &  86.22$\pm$0.54  &  6.28$\pm$0.45  &  81.94$\pm$0.35  &   8.73$\pm$0.32  \\ 
    \multicolumn{3}{c}{Wav2Vec 2.0}    & & \checkmark  &  &   89.47$\pm$0.27  &  4.58$\pm$0.15  &  88.15$\pm$0.58  &  5.52$\pm$0.25  &  83.40$\pm$0.94  &  8.22$\pm$0.44 \\
          &&  &  &  &   \checkmark   &        90.19$\pm$0.27  &  4.29$\pm$0.12  &  88.79$\pm$0.30  &  4.99$\pm$0.16  &  84.84$\pm$0.78  &  7.05$\pm$0.11  \\
    \cmidrule(r){1-3}    \cmidrule(r){4-6}	 \cmidrule(r){7-8}   \cmidrule(r){9-10}  \cmidrule(r){11-12}
          &&  & \checkmark &  &  &            92.65$\pm$0.42  &  2.95$\pm$0.15  &  91.18$\pm$0.32  &  3.58$\pm$0.11  &  86.33$\pm$0.90  &  5.37$\pm$0.32 \\ 
    \multicolumn{3}{c}{HuBert}    & & \checkmark  &  &  \textbf{93.71$\pm$0.39}  &  \textbf{2.62$\pm$0.17}  &  92.57$\pm$0.76  &  \textbf{3.04$\pm$0.15}  &  87.66$\pm$0.33  &  4.60$\pm$0.15 \\ 
          &&  &  &  &   \checkmark   &        93.64$\pm$0.43  &  2.82$\pm$0.18  &  \textbf{92.63$\pm$0.38}  &  3.23$\pm$0.19  & \textbf{88.86$\pm$0.57}  &  \textbf{4.30$\pm$0.23}  \\     
    \midrule       
   \midrule       
    \multicolumn{12}{c}{\textbf{(b) Top-2 ACC(\%) and EER(\%) with 2-mix test}} \\
    \midrule    
    \multicolumn{3}{c}{Pre-train} & \multicolumn{3}{c}{Fine-tune} & \multicolumn{2}{c}{50-shot} & \multicolumn{2}{c}{30-shot} & \multicolumn{2}{c}{15-shot}   \\
    \cmidrule(r){1-3}    \cmidrule(r){4-6}	 \cmidrule(r){7-8}   \cmidrule(r){9-10}  \cmidrule(r){11-12}
                 Clean & Mixup & MT     & Clean & Mixup & MT  & Top-2 ACC ($\uparrow$) & EER ($\downarrow$)  & Top-2 ACC ($\uparrow$)  & EER ($\downarrow$)  & Top-2 ACC ($\uparrow$) & EER ($\downarrow$)  \\
    \cmidrule(r){1-3}    \cmidrule(r){4-6}	 \cmidrule(r){7-8}   \cmidrule(r){9-10}  \cmidrule(r){11-12}
                \checkmark &&& \checkmark &&&  44.23$\pm$0.96  & 26.28$\pm$0.43  & 43.97$\pm$1.02  & 26.42$\pm$0.80  & 42.85$\pm$1.60  & 26.77$\pm$0.75 \\
                \checkmark&&&&\checkmark&&     51.98$\pm$0.32  & 21.16$\pm$0.25  & 50.00$\pm$0.79  & 22.15$\pm$0.77  & 47.32$\pm$0.75  & 23.50$\pm$0.37 \\
                \checkmark&&&&&\checkmark&     54.87$\pm$0.44  & 19.36$\pm$0.16  & 52.75$\pm$0.48  & 20.37$\pm$0.23  & 49.84$\pm$0.84  & 21.78$\pm$0.63 \\
\cmidrule(r){1-3}    \cmidrule(r){4-6}   \cmidrule(r){7-8}   \cmidrule(r){9-10}  \cmidrule(r){11-12}
                &\checkmark&& \checkmark &&&   52.67$\pm$1.17  & 23.68$\pm$0.51  & 51.33$\pm$0.95  & 24.30$\pm$0.68  & 50.19$\pm$0.96  & 25.47$\pm$0.51 \\
                &\checkmark&&&\checkmark&&     58.83$\pm$0.38  & 18.98$\pm$0.27  & 56.74$\pm$0.68  & 19.97$\pm$0.42  & 54.06$\pm$0.69  & 21.65$\pm$0.41 \\
                &\checkmark&&&&\checkmark&     62.07$\pm$0.40  & 16.65$\pm$0.12  & 60.49$\pm$0.39  & 17.23$\pm$0.15  & 56.77$\pm$0.99  & 19.11$\pm$0.50 \\
\cmidrule(r){1-3}    \cmidrule(r){4-6}   \cmidrule(r){7-8}   \cmidrule(r){9-10}  \cmidrule(r){11-12}
                &&\checkmark& \checkmark &&&   54.33$\pm$0.52  & 19.79$\pm$0.49  & 51.93$\pm$0.84  & 20.88$\pm$0.38  & 49.70$\pm$1.25  & 21.59$\pm$0.41 \\
                &&\checkmark&&\checkmark&&     60.21$\pm$0.46  & 15.98$\pm$0.34  & 57.81$\pm$0.36  & 17.15$\pm$0.17  & 53.47$\pm$1.31  & 18.80$\pm$0.57 \\
                &&\checkmark&&&\checkmark&     \textbf{64.02$\pm$0.53}  & \textbf{14.55$\pm$0.43}  & \textbf{61.02$\pm$0.54}  & \textbf{15.99$\pm$0.32}  & \textbf{56.06$\pm$1.22}  & \textbf{17.92$\pm$0.64} \\
\cmidrule(r){1-3}    \cmidrule(r){4-6}   \cmidrule(r){7-8}   \cmidrule(r){9-10}  \cmidrule(r){11-12}
          &&  & \checkmark &  &  &            35.32$\pm$1.42  & 28.80$\pm$1.02  & 31.46$\pm$1.59  & 30.38$\pm$1.32  & 27.77$\pm$1.65  & 31.46$\pm$1.68 \\
    \multicolumn{3}{c}{Wav2Vec 2.0}    & & \checkmark  &  &   45.04$\pm$0.29  & 24.06$\pm$0.30  & 41.09$\pm$0.87  & 25.59$\pm$0.56  & 34.90$\pm$1.11  & 28.19$\pm$0.79 \\
          &&  &  &  &   \checkmark   &        48.89$\pm$0.43  & 22.39$\pm$0.26  & 44.41$\pm$0.39  & 23.83$\pm$0.38  & 38.09$\pm$1.18  & 26.67$\pm$0.41  \\
    \cmidrule(r){1-3}    \cmidrule(r){4-6}	 \cmidrule(r){7-8}   \cmidrule(r){9-10}  \cmidrule(r){11-12}
          &&  & \checkmark &  &  &            53.39$\pm$0.91  & 20.80$\pm$0.65  & 50.68$\pm$0.78  & 21.66$\pm$0.45  & 47.12$\pm$0.85  & 23.48$\pm$1.00  \\
    \multicolumn{3}{c}{HuBert}    & & \checkmark  &  &  61.80$\pm$0.37  & 16.02$\pm$0.31  & 58.64$\pm$0.37  & 17.10$\pm$0.20  & 52.72$\pm$0.67  & 19.52$\pm$0.46  \\
          &&  &  &  &   \checkmark   &        \textbf{65.35$\pm$0.34}  & \textbf{14.49$\pm$0.26}  & \textbf{62.07$\pm$0.63}  & \textbf{15.48$\pm$0.15}  & \textbf{56.17$\pm$0.30}  & \textbf{17.61$\pm$0.26}  \\
    \bottomrule
    \end{tabular}}
    \vspace{-2.5mm}
\end{table*}

\subsection{Pre-training results}

We first compare the performance of the pre-trained models obtained with different training strategies, using the clean and 2-mix test sets from the pre-training dataset. The performance is evaluated by two metrics: 
(1) Equal Error Rate (EER), which measures the accuracy of the model on keyword detection, 
and (2) Top-k Accuracy, which assesses the accuracy of the model in discriminating among the keywords. Both measures can be easily computed from the detection scores produced by the classifier layer of the pre-trained model. Note that for the 2-mix test, a test trial succeeds only if the two keywords in the mixed speech are both detected and, therefore evaluated by Top-2 Accuracy. The experimental results are shown in Table~\ref{tab:compare}. 

It can be seen that in the clean speech test, Mixup marginally surpasses MT; 
however, in the mixed speech test, MT gains a significant advantage. 
This validates our hypothesis: leveraging a vast amount of mixed speech, 
MT empowers the model by allowing it to discover and present clean speech patterns from mixed signals. 

\subsection{Few-shot results}

The few-shot KWS results with different pre-training models and fine-tuning strategies are reported in Table~\ref{tab:result}. First of all, we compare the three models pre-trained by ourselves. It can be seen that the Mixup model shows the best performance in the clean test, but in the 2-mix test, the MT model is much better. This is not surprising, as MT tunes the model to tackle mixed speech, especially with limited model capacities (\emph{$\sim$4M parameters} in our case). The second consistent observation is that no matter which pre-trained model is used, MT fine-tuning shows a clear advantage compared to clean speech fine-tuning and Mixup fine-tuning. This conclusion holds for both the clean test and the 2-mix test, but the advantage is more evident in the 2-mix test. This indicates that MT is a vital technology for few-shot KWS. It should be always used in the fine-tuning stage. However, whether it should used in the pre-training phase depends on the test condition: if the test condition involves a significant proportion of mixed speech, it should be definitely used; otherwise Mixup is likely a better pre-training strategy. Once again, this conclusion is based on our present experimental settings. It is still unknown if MT is more effective for pre-training with a larger model or with more data, and whether combining Mixup and MT can offer better pre-training is also under investigation.

The results with SSL models are even more significant. The first observation is that with both the Wav2Vec2.0 model and the HuBert model, excellent performance is obtained on the clean test, and the performance gap is margin with all three fine-tuning strategies. This is fully expected as the two SSL models are larger (\emph{$\sim$94M parameters} for Wav2Vec2.0 and HuBert) and were trained with much more speech data than our models. In the 2-mix test, however, the performance is very different: the Wav2Vec2.0 model largely failed while the HuBert model yielded strong results. This indicates some quite different properties inside the two SSL models that require further investigation. Nevertheless, MT is an important ingredient for the good performance obtained with the HuBert model. 
Overall, HuBert + MT forms a powerful solution that can achieve universally good results in all the test conditions. Interestingly, the MT + MT strategy offers quite close performance to HuBert + MT, although the MT pre-trained model is much smaller and was trained with less data. This suggests that MT pre-training holds much potential and deserves further investigation.

\section{Conclusions}

This paper focuses on few-shot keyword spotting from mixed speech, within the pre-training and fine-tuning framework. Three training strategies, the conventional clean speech training, Mixup training, and recently proposed Mix Training (MT) were employed to perform pre-training and/or fine-tuning. Empirical results show that for the clean test, Mixup + MT yields the best performance, though, in the mixed speech test, MT + MT is the most powerful approach. This provides a strong evidence that MT is a vital technology for dealing with mixed speech. Moreover, by combining a large-scale HuBert model with MT, we obtained very strong performance in both the clean and mixed speech tests. In the future work, we will investigate the MT pre-training with a large model structure, and most interestingly, MT based on self-supervised learning. 



\newpage

\bibliographystyle{IEEEtran}
\bibliography{my_reference}

\begin{thebibliography}{10}
\providecommand{\url}[1]{#1}
\csname url@samestyle\endcsname
\providecommand{\newblock}{\relax}
\providecommand{\bibinfo}[2]{#2}
\providecommand{\BIBentrySTDinterwordspacing}{\spaceskip=0pt\relax}
\providecommand{\BIBentryALTinterwordstretchfactor}{4}
\providecommand{\BIBentryALTinterwordspacing}{\spaceskip=\fontdimen2\font plus
\BIBentryALTinterwordstretchfactor\fontdimen3\font minus
  \fontdimen4\font\relax}
\providecommand{\BIBforeignlanguage}[2]{{%
\expandafter\ifx\csname l@#1\endcsname\relax
\typeout{** WARNING: IEEEtran.bst: No hyphenation pattern has been}%
\typeout{** loaded for the language `#1'. Using the pattern for}%
\typeout{** the default language instead.}%
\else
\language=\csname l@#1\endcsname
\fi
#2}}
\providecommand{\BIBdecl}{\relax}
\BIBdecl

\bibitem{teacher1967experimental}
C.~Teacher, H.~Kellett, and L.~Focht, ``Experimental, limited vocabulary,
  speech recognizer,'' \emph{IEEE Transactions on Audio and Electroacoustics},
  vol.~15, no.~3, pp. 127--130, 1967.

\bibitem{he2017streaming}
Y.~He, R.~Prabhavalkar, K.~Rao, W.~Li, A.~Bakhtin, and I.~McGraw, ``Streaming
  small-footprint keyword spotting using sequence-to-sequence models,'' in
  \emph{ASRU}.\hskip 1em plus 0.5em minus 0.4em\relax IEEE, 2017, pp. 474--481.

\bibitem{zhao2020end}
Z.~Zhao and W.~Zhang, ``End-to-end keyword search based on attention and energy
  scorer for low resource languages.'' in \emph{INTERSPEECH}, 2020, pp.
  2587--2591.

\bibitem{zhang2021efficient}
S.~Zhang, T.~Zhang, S.~Chen, F.~Chen, and X.~Yin, ``An efficient temporal model
  for small-footprint keyword spotting,'' in \emph{IC-NIDC}.\hskip 1em plus
  0.5em minus 0.4em\relax IEEE, 2021, pp. 289--293.

\bibitem{liu2022neural}
Z.~Liu, T.~Li, and P.~Zhang, ``Neural keyword confidence estimation for
  open-vocabulary keyword spotting,'' \emph{Electronics Letters}, vol.~58,
  no.~3, pp. 133--135, 2022.

\bibitem{lee2023phonmatchnet}
Y.-H. Lee and N.~Cho, ``Phonmatchnet: phoneme-guided zero-shot keyword spotting
  for user-defined keywords,'' \emph{arXiv preprint arXiv:2308.16511}, 2023.

\bibitem{hoy2018alexa}
M.~B. Hoy, ``Alexa, siri, cortana, and more: an introduction to voice
  assistants,'' \emph{Medical reference services quarterly}, vol.~37, no.~1,
  pp. 81--88, 2018.

\bibitem{van2022feature}
E.~van~der Westhuizen, H.~Kamper, R.~Menon, J.~Quinn, and T.~Niesler, ``Feature
  learning for efficient asr-free keyword spotting in low-resource languages,''
  \emph{Computer Speech \& Language}, vol.~71, p. 101275, 2022.

\bibitem{gao2020towards}
Y.~Gao, Y.~Mishchenko, A.~Shah, S.~Matsoukas, and S.~Vitaladevuni, ``Towards
  data-efficient modeling for wake word spotting,'' in \emph{ICASSP}.\hskip 1em
  plus 0.5em minus 0.4em\relax IEEE, 2020, pp. 7479--7483.

\bibitem{majumdar2020matchboxnet}
S.~Majumdar and B.~Ginsburg, ``Matchboxnet: 1d time-channel separable
  convolutional neural network architecture for speech commands recognition,''
  \emph{arXiv preprint arXiv:2004.08531}, 2020.

\bibitem{chen2018investigation}
Y.~Chen, T.~Ko, L.~Shang, X.~Chen, X.~Jiang, and Q.~Li, ``An investigation of
  few-shot learning in spoken term classification,'' \emph{arXiv preprint
  arXiv:1812.10233}, 2018.

\bibitem{chen2021meta}
Y.~Chen, T.~Ko, and J.~Wang, ``A meta-learning approach for user-defined spoken
  term classification with varying classes and examples.'' in
  \emph{Interspeech}, 2021, pp. 4224--4228.

\bibitem{parnami2022few}
A.~Parnami and M.~Lee, ``Few-shot keyword spotting with prototypical
  networks,'' in \emph{ICMLT}, 2022, pp. 277--283.

\bibitem{lin2020training}
J.~Lin, K.~Kilgour, D.~Roblek, and M.~Sharifi, ``Training keyword spotters with
  limited and synthesized speech data,'' in \emph{ICASSP}.\hskip 1em plus 0.5em
  minus 0.4em\relax IEEE, 2020, pp. 7474--7478.

\bibitem{awasthi2021teaching}
A.~Awasthi, K.~Kilgour, and H.~Rom, ``Teaching keyword spotters to spot new
  keywords with limited examples,'' \emph{arXiv preprint arXiv:2106.02443},
  2021.

\bibitem{mazumder2021few}
M.~Mazumder, C.~Banbury, J.~Meyer, P.~Warden, and V.~J. Reddi, ``Few-shot
  keyword spotting in any language,'' \emph{arXiv preprint arXiv:2104.01454},
  2021.

\bibitem{jung2023metric}
J.~Jung, Y.~Kim, J.~Park, Y.~Lim, B.-Y. Kim, Y.~Jang, and J.~S. Chung, ``Metric
  learning for user-defined keyword spotting,'' in \emph{ICASSP}.\hskip 1em
  plus 0.5em minus 0.4em\relax IEEE, 2023, pp. 1--5.

\bibitem{rusci2023few}
M.~Rusci and T.~Tuytelaars, ``Few-shot open-set learning for on-device
  customization of keyword spotting systems,'' \emph{arXiv preprint
  arXiv:2306.02161}, 2023.

\bibitem{seo2021wav2kws}
D.~Seo, H.-S. Oh, and Y.~Jung, ``Wav2kws: Transfer learning from speech
  representations for keyword spotting,'' \emph{IEEE Access}, vol.~9, pp.
  80\,682--80\,691, 2021.

\bibitem{kao2023efficiency}
W.-T. Kao, Y.-K. Wu, C.-P. Chen, Z.-S. Chen, Y.-P. Tsai, and H.-Y. Lee, ``On
  the efficiency of integrating self-supervised learning and meta-learning for
  user-defined few-shot keyword spotting,'' in \emph{SLT}.\hskip 1em plus 0.5em
  minus 0.4em\relax IEEE, 2023, pp. 414--421.

\bibitem{yang2023improving}
S.~Yang, B.~Kim, K.~Shim, and S.~Chang, ``Improving small footprint few-shot
  keyword spotting with supervision on auxiliary data,'' \emph{arXiv preprint
  arXiv:2309.00647}, 2023.

\bibitem{seltzer2012acoustic}
M.~L. Seltzer, ``Acoustic model training for robust speech recognition,''
  \emph{Techniques for Noise Robustness in Automatic Speech Recognition}, pp.
  347--368, 2012.

\bibitem{raju2018data}
A.~Raju, S.~Panchapagesan, X.~Liu, A.~Mandal, and N.~Strom, ``Data augmentation
  for robust keyword spotting under playback interference,'' \emph{arXiv
  preprint arXiv:1808.00563}, 2018.

\bibitem{park2021noisy}
H.-J. Park, P.~Zhu, I.~L. Moreno, and N.~Subrahmanya, ``Noisy student-teacher
  training for robust keyword spotting,'' \emph{arXiv preprint
  arXiv:2106.01604}, 2021.

\bibitem{zhang2017mixup}
H.~Zhang, M.~Cisse, Y.~N. Dauphin, and D.~Lopez-Paz, ``mixup: Beyond empirical
  risk minimization,'' \emph{arXiv preprint arXiv:1710.09412}, 2017.

\bibitem{shi2023spot}
Y.~Shi, D.~Wang, L.~Li, J.~Han, and S.~Yin, ``Spot keywords from very noisy and
  mixed speech,'' \emph{arXiv preprint arXiv:2305.17706}, 2023.

\bibitem{snell2017prototypical}
J.~Snell, K.~Swersky, and R.~Zemel, ``Prototypical networks for few-shot
  learning,'' \emph{Advances in neural information processing systems},
  vol.~30, 2017.

\bibitem{yangrobust}
C.~Yang, Y.~M. Saidutta, R.~S. Srinivasa, C.-H. Lee, Y.~Shen, and H.~Jin,
  ``Robust keyword spotting for noisy environments by leveraging speech
  enhancement and speech presence probability,'' in \emph{Interspeech}, 2013,
  pp. 1638--1642.

\bibitem{bonet2021speech}
D.~Bonet, G.~C{\'a}mbara, F.~L{\'o}pez, P.~G{\'o}mez, C.~Segura, and J.~Luque,
  ``Speech enhancement for wake-up-word detection in voice assistants,''
  \emph{arXiv preprint arXiv:2101.12732}, 2021.

\bibitem{verma2019manifold}
V.~Verma, A.~Lamb, C.~Beckham, A.~Najafi, I.~Mitliagkas, D.~Lopez-Paz, and
  Y.~Bengio, ``Manifold mixup: Better representations by interpolating hidden
  states,'' in \emph{ICML}.\hskip 1em plus 0.5em minus 0.4em\relax PMLR, 2019,
  pp. 6438--6447.

\bibitem{yun2019cutmix}
S.~Yun, D.~Han, S.~J. Oh, S.~Chun, J.~Choe, and Y.~Yoo, ``Cutmix:
  Regularization strategy to train strong classifiers with localizable
  features,'' in \emph{IEEE/CVF ICCV}, 2019, pp. 6023--6032.

\bibitem{hendrycks2022pixmix}
D.~Hendrycks, A.~Zou, M.~Mazeika, L.~Tang, B.~Li, D.~Song, and J.~Steinhardt,
  ``Pixmix: Dreamlike pictures comprehensively improve safety measures,'' in
  \emph{IEEE/CVF CVPR}, 2022, pp. 16\,783--16\,792.

\bibitem{kim2021co}
J.-H. Kim, W.~Choo, H.~Jeong, and H.~O. Song, ``Co-mixup: Saliency guided joint
  mixup with supermodular diversity,'' \emph{arXiv preprint arXiv:2102.03065},
  2021.

\bibitem{tripathi2020end}
A.~Tripathi, H.~Lu, and H.~Sak, ``End-to-end multi-talker overlapping speech
  recognition,'' in \emph{ICASSP 2020-2020 IEEE International Conference on
  Acoustics, Speech and Signal Processing (ICASSP)}.\hskip 1em plus 0.5em minus
  0.4em\relax IEEE, 2020, pp. 6129--6133.

\bibitem{mcauliffe2017montreal}
M.~McAuliffe, M.~Socolof, S.~Mihuc, M.~Wagner, and M.~Sonderegger, ``Montreal
  forced aligner: Trainable text-speech alignment using kaldi.'' in
  \emph{INTERSPEECH}, vol. 2017, 2017, pp. 498--502.

\bibitem{warden2018speech}
P.~Warden, ``Speech commands: A dataset for limited-vocabulary speech
  recognition,'' \emph{arXiv preprint arXiv:1804.03209}, 2018.

\bibitem{tan2019efficientnet}
M.~Tan and Q.~Le, ``Efficientnet: Rethinking model scaling for convolutional
  neural networks,'' in \emph{ICML}.\hskip 1em plus 0.5em minus 0.4em\relax
  PMLR, 2019, pp. 6105--6114.

\bibitem{zou2023mixupe}
Y.~Zou, V.~Verma, S.~Mittal, W.~H. Tang, H.~Pham, J.~Kannala, Y.~Bengio,
  A.~Solin, and K.~Kawaguchi, ``Mixupe: Understanding and improving mixup from
  directional derivative perspective,'' in \emph{Uncertainty in Artificial
  Intelligence}.\hskip 1em plus 0.5em minus 0.4em\relax PMLR, 2023, pp.
  2597--2607.

\bibitem{baevski2020wav2vec}
A.~Baevski, Y.~Zhou, A.~Mohamed, and M.~Auli, ``wav2vec 2.0: A framework for
  self-supervised learning of speech representations,'' \emph{Advances in
  neural information processing systems}, vol.~33, pp. 12\,449--12\,460, 2020.

\bibitem{hsu2021hubert}
W.-N. Hsu, B.~Bolte, Y.-H.~H. Tsai, K.~Lakhotia, R.~Salakhutdinov, and
  A.~Mohamed, ``Hubert: Self-supervised speech representation learning by
  masked prediction of hidden units,'' \emph{IEEE/ACM Transactions on Audio,
  Speech, and Language Processing}, vol.~29, pp. 3451--3460, 2021.

\end{thebibliography}

\end{document}